\documentclass[pra,showpacs]{revtex4}
\usepackage{amssymb}
\usepackage{graphicx}
\usepackage[dvipdfm]{hyperref}
\usepackage{verbatim}
\usepackage[normalem]{ulem}
\usepackage{color}
\usepackage{amsmath}

\begin{document}

\title{Quantum versus mean-field collapse in a many-body system.}
\author{G. E. Astrakharchik$^{1}$}
\author{B. A. Malomed$^{2}$}
\affiliation{$^{1}$Departament de F\'{\i}sica i Enginyeria Nuclear, Universitat Polit\`{e}cnica de Catalunya, Barcelona, Spain}
\affiliation{$^{2}$Department of Physical Electronics, School of Electrical Engineering,
Faculty of Engineering, Tel Aviv University, Tel Aviv University, Tel Aviv 69978, Israel}

\begin{abstract}
The recent analysis, based on the mean-field approximation (MFA), has predicted that the critical quantum collapse of the bosonic wave function, pulled to the center by the inverse-square potential in the three-dimensional space, is suppressed by the repulsive cubic nonlinearity in the bosonic gas, the collapsing ground state being replaced by a regular one.
We demonstrate that a similar stabilization acts in a quantum many-body system, beyond the MFA.
While the collapse remains possible, repulsive two-particle interactions give rise to a metastable gaseous state, which is separated by a potential barrier from the collapsing regime.
The stability of this state improves with the increase of the number of particles.
The results are produced by calculations of the variational energy, with the help of the Monte Carlo method.
\end{abstract}

\date{August 21, 2015}
\pacs{03.75.Hh; 03.75.Kk, 31.15.-p; 03.75.Lm}
\maketitle

\section{Introduction}

It is well known that the attractive inverse-square potential,
\begin{equation}
U(r)=-U_{0}r^{-2},  \label{pot}
\end{equation}
causes the quantum collapse (alias \textquotedblleft fall onto the center\textquotedblright ) of a wave function of the nonrelativistic particle in three dimensions (3D).
Potential~(\ref{pot}) drives the quantum collapse in its critical form~\cite{LL}: if $U_{0}$ exceeds a critical value [which is $1/8$ in the present notation, see Eq.~(\ref{GPE}) below], the corresponding Schr\"{o}dinger equation fails to produce the ground state (GS) (or, loosely speaking, in this case the energy of the GS falls to $E\rightarrow -\infty $, while the GS radius collapses to zero).
The critical form of the quantum collapse is explained by the fact that the quantization breaks the scaling invariance of the classical mechanical system with the same potential~(\ref{pot}).
The effect is also named \textit{quantum anomaly}, or \textquotedblleft \textit{dimensional transmutation}\textquotedblright ~\cite{anomaly}.

The first solution of the problem of the missing GS was proposed in terms of the linear quantum field theory (rather than quantum mechanics, where the problem arises)~\cite{anomaly}.
That approach actually postulated the existence of the GS, with an \emph{arbitrary} spatial size, which was used as a basic scale for the application of the quantum-field renormalization procedure.

A different solution of the quantum-collapse problem was proposed Refs.~\cite{HS} and~\cite{HS2}, which introduced, in lieu of the single quantum particle, a bosonic gas, pulled to the center by potential~(\ref{pot}), and described, in the framework of the mean-field approximation (MFA), by the
Gross-Pitaevskii equation (GPE) for the single-particle wave function~\cite{HS}:
\begin{equation}
i\hbar \frac{\partial \psi }{\partial t}=\frac{\hbar ^{2}}{m}\left( -\frac{1}{2}\nabla ^{2}\psi -\frac{U_{0}}{r^{2}}\psi +4\pi a_s|\psi |^{2}\psi \right) ,  \label{GPE}
\end{equation}
where $m$ is the particle's mass. A key ingredient which secures the GS against the collapse in the framework of the GPE is the self-repulsive nonlinear term in Eq.~(\ref{GPE}).
The strength of the self-repulsion is determined by the $s$-wave scattering length, $a_s$, of two-particle collisions~\cite{Pitaevskii}.
Similarly, the collisional stabilization of a two-component gas pulled to the center by potential~(\ref{pot}) is predicted by the system of two coupled GPEs~\cite{HS2}.

The physical realization of the setting, which was proposed in Ref.~\cite{HS}, assumes an ultra-cold gas of molecules carrying a permanent electric dipole moment, $d$ (LiCs~\cite{LiCs} and KRb~\cite{KRb} molecules may be appropriate for this purpose), which are pulled to a point-like electric charge $Q$ placed at the central position (in particular, it may be an ion held by an optical trapping potential~\cite{ion}).
In this case, the strength of potential~(\ref{pot}) is $U_{0}=|Q|d$.
An alternative to the molecular gas may be provided by a gas of atoms kept in a long-lived Rydberg
state~\cite{Rydberg-review}, which also carry a dipole electric moment.
The long-range dipole-dipole interactions in the same setting were also taken into account in the mean-field theory elaborated in Refs.~\cite{HS} and~\cite{HS2}.
This was done assuming that an individual dipole moment interacts with the mean electrostatic field, created by all other particles, i.e., the GPE was augmented by the term accounting for the coupling of the dipole moment to the local electrostatic field, and supplemented by the respective Poisson equation. As a result, the strength of the contact repulsive nonlinearity was subject to an effective renormalization, with an addition induced by the dipole-dipole coefficients: $a_{s}\rightarrow
a_{s}+md^{2}/\hbar ^{2}$.

Thus, it has been demonstrated in Refs.~\cite{HS} and~\cite{HS2} that the repulsive nonlinearity suppresses the quantum collapse, creating the missing GS at all values of $U_{0}$, both above and below the above-mentioned critical one, $U_{0}=1/8$.
An estimate, based on relevant values of the underlying physical parameters, predicted the radius of the restored GS to be on the order of a few microns.
At $U_{0}>1/8$, a phase transition in the newly built GS was found at $U_{0}=1/2$.
At that point, the GPE features a breakup of the analyticity in the GS wave function, and a change in the structure of its wave function at $r\rightarrow 0$ (however, the GS survives the transition).
This quantum phase transition resembles those found in other systems~\cite{Grisha}.

In addition to Eq.~(\ref{GPE}) with the isotropic attractive potential~(\ref{pot}), its version with the symmetry reduced from spherical to cylindrical, which corresponds to potential $U_{0}r^{-2}\cos \theta $, where $\theta $ is the angular coordinate in the 3D space, was elaborated in Ref.~\cite{HS-cyl}.
Furthermore, in Ref.~\cite{HS}, the same problem was considered in the two-dimensional (2D) geometry with attractive potential~(\ref{pot}), for which quantum mechanics gives rise to the collapse for any $U_{0}>0$.
Aside from the aforementioned realizations, the 2D system models a gas of polarizable atoms that do not carry a permanent electric moment, while an effective moment is induced in them by the electric field of a uniformly charged wire piercing the system's plane~\cite{Schmiedmayer}.
Similarly, a magnetic moment may be induced by a current filament set perpendicular to the plane~\cite{HS}.
In either case, the 2D version of potential~(\ref{pot}) is generated by the attraction of the induced moment to the source which induces it.
The difference from the 3D setting is that the cubic nonlinearity does not suppress the quantum collapse in the 2D setting, only the quintic term (if it is a physically relevant one) being sufficiently
strong for that~\cite{HS}.

The results demonstrating the stabilization of the bosonic gases, against the fall onto the center, by repulsive interactions were obtained in Refs.~\cite{HS}, \cite{HS2}, and~\cite{HS-cyl} by means of the MFA.
The objective of the present work is to address a natural problem suggested by those results, namely, whether similar effects can be found beyond the limits of the MFA, i.e., in the framework of a direct analysis of a system of quantum particles pulled to the center by potential~(\ref{pot}).
The corresponding many-body model is formulated in Section~\ref{sec:II}.
As a preliminary step, the analysis of the single-particle situation is carried out, using the variational approach, in Section~\ref{sec:III}.
The mean-field study of the many-body system based on the local-density approximation is carried out
in Section~\ref{sec:LDA}.
Physically relevant interparticle interaction potentials, namely, of the hard- and soft-core types, are introduced in Section~\ref{sec:IV}.
The Monte Carlo method, used for calculations of the variational energy, is outlined in Section~\ref{sec:V}.
The most essential results, obtained by means of this method, are summarized in Section~\ref{sec:VI}, for the many-body system composed of 2 to 10000 interacting particles.
The consideration of a relatively small number of particles, $N$, is essential, as the MFA should be definitely valid for $N\gg 1$, in which case the direct analysis of the interactions between particles may only produce post-MFA effects, such as two-body correlations.
We infer that, while the collapse remains possible in the many-body system, the repulsive interactions between particles give rise to a metastable gaseous state, separated from the collapsing GS by a sufficiently tall energy barrier.
Thus, the possibility of the collisional stabilization of the quantum gas attracted to the center by potential~(\ref{pot}), predicted by the MFA, is justified by the many-body theory.
The paper is concluded by Section~\ref{sec:VII}.
In Appendix~\ref{sec:Appendix} we derive the virial theorem for particles interacting via zero-range pseudopotential.

\section{The Hamiltonian\label{sec:II}}

We study a system of $N$ particles with repulsive interactions, in the presence of the inverse-square potential~(\ref{pot}), which pulls all the particles to the center.
As said above, the system may be unstable against the quantum collapse, alias fall onto the center, while the MFA, which takes into regard both the contact and long-range two-particle interactions, predicts a possibility of the stabilization~\cite{HS}.
In terms of the quantum-mechanical description, the many-body Hamiltonian is
\begin{equation}
\hat{H}
= -\sum\limits_{j=1}^{N}\left(
\frac{\hbar^2\nabla_{j}^{2}}{2m}
+
\frac{U_{0}}{r_{j}^{2}}
\right) +\sum\limits_{j<k}^{N}V_{\mathrm{int}}(|\mathbf{r}_{j}-\mathbf{r}_{k}|),  \label{H}
\end{equation}
where $U_{0}$ is the strength of central potential~(\ref{pot}), $\mathbf{r}_{j}$ are coordinates of the $j$-th particle in the 3D space, $m$ is, as in Eq.~(\ref{GPE}), the particle's mass, and $V_{\mathrm{int}}(r)$ is the repulsive interaction potential between the particles.
In the MFA, $V_{\mathrm{int}}(r)$ is characterized solely by the $s$-wave scattering length~\cite{Pitaevskii}, plus the dipole moment which determines the additional mean-field interaction of the given particle with the electrostatic field created by all others~\cite{HS}.
Two basic forms of the interaction potential chosen for the analysis in the present work are specified below.

As mentioned above, the attractive potential~(\ref{pot}) is a peculiar one, as it scales with distance $r$ in exactly the same way as the Laplacian operator of the kinetic energy; as a result, the free-particle problem, with $V_{\mathrm{int}}(|\mathbf{r}_{j}-\mathbf{r}_{k}|)=0$, has no intrinsic
length scale.
In other words, the system's properties scale, following variation of the particle density, making phase transitions impossible.
This feature is in a stark contrast with the usual situation, where a phase transition may be reached by altering the density.
The absence of a characteristic length implies that the state of the system is fully defined by the attraction strength, $U_{0}$, and there is the above-mentioned critical value, $U_{0}=1/8$, above which the collapse occurs in the quantum-mechanical system~\cite{LL}.
The terms in Hamiltonian~(\ref{H}) which account for the attraction to the center and kinetic energy scale linearly with the number of particles, $N$, while the energy of the repulsive interparticle interactions scales as $N^{2}$.
This fact suggests the possibility of the stabilization against the collapse in the framework
of the many-body theory.

One may foresee two opposite scenarios related to the possibility of the collapse in the system.
In the collapsed state, particles fall onto the center, $r_{j}\rightarrow 0$, creating a diverging density, hence the total energy is negative and diverges too.
On the other side, in a gaseous state the particles are not bound to the center and fly away, $r_{j}\rightarrow \infty $, leading to an asymptotically zero density and zero energy.
To test what scenario is energetically preferable, we will add an artificial Gaussian factor to the wave function, which will localize particles within a finite distance from the center.

Thus, the single-particle wave function is chosen in the form of a product of a Gaussian and a power-law radial term,
\begin{equation}
f_{1}(r)=r^{\beta }\exp (-\alpha r^{2}),  \label{Eq:f1}
\end{equation}
where the inverse Gaussian width, $\alpha \geq 0$, determines the \textit{inverse localization length}.
It affects the system's size and, consequently, the density.
Alternatively, parameter $\alpha$ can be interpreted in terms of an effective external harmonic confinement with frequency $\omega =2\alpha \hbar /m$.
On the other hand, the shape of the cloud at $r_{j}\rightarrow 0$ is controlled by parameter $\beta $ in ansatz~(\ref{Eq:f1}).

\section{The single-particle solution\label{sec:III}}

\begin{figure}[tbp]
\includegraphics[width=0.4\columnwidth, angle=0]{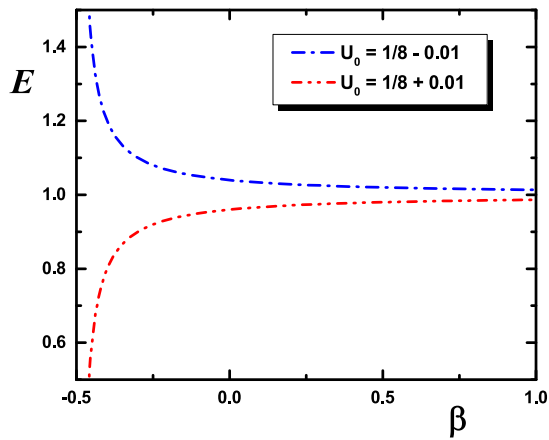}
\caption{(Color online) Variational energy~(\ref{Eq:E1}) for a single particle with $\alpha =1$ versus $\beta$ for values of the strength of the central attractive potential close to the critical value, $U_{0}=1/8$. The upper and lower curves correspond, severally, to free state at $U_{0}<1/8$, and a collapsed one at $U_{0}>1/8$. }
\label{FigEN1}
\end{figure}

The single-particle problem, defined by Hamiltonian~(\ref{H}) with $N=1$, can be studied by means of the variational method, treating $\alpha$ and $\beta$ in ansatz~(\ref{Eq:f1}) as variational parameters.
In the single-particle sector, the system is steered by the competition of the external potential and kinetic energy, while the interparticle potential, $V_{\mathrm{int}}(|\mathbf{r}_{i}-\mathbf{r}_{j}|)$, does not appear.
The variational energy,
$E_{\mathrm{var}}^{(1)}=\left[ \int f_{1}^{2}(\mathbf{r})\;d\mathbf{r}\right] ^{-1}\int f_{1}(\mathbf{r})Hf_{1}(\mathbf{r})d\mathbf{r}$, with $f_{1}$ taken as per Eq.~(\ref{Eq:f1}), can be evaluated explicitly:
\begin{equation}
E^{(1)}=\alpha \left[ 1-\frac{8U_{0}-1}{2(1+2\beta )}\right] .  \label{Eq:E1}
\end{equation}
For a fixed localization size, $\alpha =\mathrm{const}$, this energy is a decreasing function of $\beta$ if $U_{0}$ is smaller than the critical value, $U_{0}=1/8$ (the same which was mentioned above), and an increasing function for $U_{0}>1/8$, see Fig.~\ref{FigEN1}.
By varying $\beta$, we probe how the energy responds to the increase or decrease of the density at
the center of the cloud.

For $U_{0}<1/8$, the minimum of energy~(\ref{Eq:E1}) with respect to $\beta$ is $E=\alpha $, which is reached at $\beta \rightarrow \infty$, corresponding to a delocalized state with the particles flying away from the center.
The minimum of the energy with respect to $\alpha $ is $E=0$, which is attained at $\alpha=0$, i.e., for zero density, when the localization factor in Eq.~(\ref{Eq:f1}) is absent.

The denominator in expression~(\ref{Eq:E1}) vanishes at $\beta =-1/2$, when the energy diverges.
Close to this point, at $\beta =-1/2+\epsilon $ with small $\epsilon $, the energy is $E^{(1)}\approx \left( 2\alpha /\epsilon \right) (1/8-U_{0}+\epsilon /2)$.
Depending of the sign of $1/8-U_{0}$, the energy diverges to negative or positive infinity.
Thus, for $U_{0}>1/8$, the collapsed state is realized, corresponding to $\alpha \rightarrow \infty$
(zero localization radius).
On the other hand, for $\beta \rightarrow -1/2$ in the absence of the confinement, $\alpha =0$, the density profile at small distances is $\rho (r)\propto f_{1}^{2}(r)=r^{2\beta}\equiv 1/r$, the
respective expression for the potential energy, $\int_{0}^{\infty }\rho (r)U_{0}/r^{2}\;4\pi r^{2}dr$, featuring a logarithmic divergence.

The divergence at $U_0 > 1/8$ will persist in the many-body system.
This can be seen as follows: one can place all particles but one off the center, which will cost a finite energy.
Still, at $U_{0}>1/8$, the part of the total energy corresponding to the single particle staying near the center will be divergent and negative.
That single particle falls onto the center.
One realizes that if the potential energy of the pair-interaction remains finite, all other particles will also fall onto the center.
In other words, the \emph{absolute stability threshold} remains the same, $U_{0}=1/8$, also in a many-body system.
On the other hand, a \textit{metastable} state may appear in the many-body system with repulsive interparticle interactions, as predicted by the mean-field consideration~\cite{HS}.

\section{The mean-field theory based on the local-density approximation\label{sec:LDA}}

Here we address the problem of stability of the many-body system within the framework of the local-density approximation (LDA).
This method is commonly applied to many-body systems in the presence of an external potential, when the chemical potential of the homogeneous system is known.
We use the mean-field expression for the chemical potential, $\mu _{\mathrm{\hom }}=gn$ , where $g=4\pi \hbar ^{2}a_{s}/m$ is the coupling constant.
This choice corresponds to the short-range interaction potential determined by the $s$ -wave scattering length as per the lowest Born approximation.
Mathematically, pseudopotential $V(r)=g\delta(r)$ leads to the same result.
It is important to note that this is a soft-core interaction, as it allows particles to overlap.
The pseudopotential interaction permits to derive the virial theorem in the presence of the central potential, see Eq.~(\ref{Eq:virial}) in Appendix.
It is of interest to introduce inhomogeneity by an external trap and check if the energy remains positive or negative once the trap is suddenly removed.

In the framework of the LDA, the chemical potential in the presence of an external potential is approximated by the sum of the \textit{local} chemical potential $\mu _{\mathrm{loc}}=gn$, where, this time, $n$ is a function of the coordinates, rather than a constant, and the external potential,
\begin{equation}
\mu =\mu _{\mathrm{loc}}-\frac{U_{0}}{r^{2}}+\frac{1}{2}m\omega ^{2}r^{2},
\label{Eq:LDA}
\end{equation}
where the harmonic-oscillator confinement, with the respective length scale,
$a_{\mathrm{ho}}=\sqrt{\hbar / (m\omega)}$,
is added to make the size of the system finite.
Solving Eq.~({\ref{Eq:LDA}}) for the density, one obtains the following density profile:
\begin{equation}
n(r)=\frac{1}{g}
\begin{cases}
\mu -\frac{1}{2}m\omega ^{2}r^{2}+U_{0}r^{-2}, & \mathrm{at\ }r<R_{\mathrm{TF}}, \\
0, & \mathrm{at\ }\geq R_{\mathrm{TF}},
\end{cases}
\label{Eq:LDA:n}
\end{equation}
where the cloud's radius is given by the Thomas-Fermi (TF) expression,
$R_{\mathrm{TF}} = \sqrt{\mu +\sqrt{\mu^2 + 2mU_{0}\omega^2}} / (\sqrt{m}\omega)$,
which diverges for a shallow trap, at $\omega \rightarrow 0$.
Close to $r=R_{\mathrm{TF}}$, the density profile follows a typical inverted-parabola shape of a trapped Bose condensate.
The density at the center features an integrable divergency reflecting the presence of the attractive central potential.
The value of the chemical potential itself is fixed by the normalization condition, $4\pi \int_{0}^{R_{\mathrm{TF}}}n(r)r^{2}dr=N$.

The system's energy is obtained by integrating the chemical potential: $E=\int_{0}^{N}\mu (N^{\prime })dN^{\prime }$.
We are interested to find the \emph{release energy}, which remains in the system when the harmonic
trapping is suddenly switched off, $E_{\mathrm{rel}} = E - \frac{1}{2}Nm\omega^{2}\langle r^{2}\rangle$, where $\left\langle ...\right\rangle$ stands for the average value.
It is especially important to find out if $E_{\mathrm{rel}}$ is positive or negative.

In the limit of the vanishing central potential, $U_{0}=0$, the size of the cloud is defined by the balance between the chemical potential (which depends on the particle-particle interaction strength) and the harmonic confinement~\cite{Pitaevskii}, $R_{\mathrm{TF}}=\sqrt{2\mu /m\omega^{2}}$.
The respective chemical potential, $\mu =\frac{1}{2}\hbar \omega \left(15Na_s/a_{\mathrm{ho}}\right) ^{2/5}$, is always positive, hence the respective release energy, $E_{\mathrm{rel}}=(2/7)\mu N$, is positive too.

In the opposite limit of strong attraction to the center, the chemical potential, $\mu \approx -4m^{2}U_{0}^{3}/(9N^{2}\hbar^{4}a_s^{2})$, is negative and independent of the trapping frequency, while the TF radius is inversely proportional to the strength of the attractive potential:
$R_{\mathrm{TF}}=3N\hbar^{2}a_s/(2mU_{0})$.
In this case, the system's energy diverges for on the lower integration bound corresponding to small system's sizes, $E\rightarrow -\infty$.
Thus we conclude, that, in the framework of the LDA, the collapse is always possible, as far as the energy is considered.

It is important to note that the kinetic energy, which is a key ingredient precluding the onset of the collapse, is not fully taken into account by the LDA.
A more complete mean-field Gross-Pitaevskii theory, which takes into account the kinetic energy properly, predicts the suppression of the collapse \cite{HS}. Thus, it is obviously interesting to see how beyond-mean-field effects modify that result.

\section{Soft- and hard-core interparticle potentials\label{sec:IV}}

In the single-particle problem, the collapsed state results in a density profile which is a delta-function at the center, $\rho (\mathbf{r})=\delta(\mathbf{r})$.
Then, two essential questions arise:
(i) whether a similar singular state, $\rho (\mathbf{r})=N\delta (\mathbf{r})$, corresponds to the
GS of a many-body system containing $N$ particles; and
(ii) if a nonsingular gaseous state may be metastable, i.e., dynamically stable against small
perturbations, even if the GS corresponds to the collapse.

The answer to question (i) depends on the form of the interparticle potential, $V_{\mathrm{int}}(r)$, at small $r$.
If the interaction features a \emph{soft core}, i.e., there is an upper bound on the interparticle
interaction potential $V_{0}$, such that $V(r)<V_{0}$ at $r\rightarrow 0$, then there also exists an upper bound on the particle-particle interaction energy: $E_{\mathrm{int}}<V_{0}N(N-1)/2$.
At the same time, the potential energy corresponding to the central attractive potential diverges to $-\infty$ at $U_{0}>1/8$.
As a result, any soft-core potential is not sufficient to prevent the system from the transition to the collapsed GS (fall onto the center).
On the other hand, for physically realistic interparticle potentials with a hard core (for example, of the Van der Waals type), with $V_{0}\rightarrow \infty $, the fully-collapsed state, with $\rho (\mathbf{r})=N\delta (\mathbf{r})$, would lead to a diverging positive contribution to the energy, which may prevent this state from being the GS.

The answer to question (ii) depends on the number of particles, $N$, and parameters of the interaction. It may happen that, while the GS corresponds to the fall onto the center, there is a barrier between the collapsed and nonsingular gaseous states, making the latter one metastable.
Indeed, the interparticle-interaction energy, $E_{\mathrm{int}}$, depends quadratically on $N$, while the energy of the attraction to the central potential scales linearly with $N$, which suggests that the systems may effectively stabilize itself for $N$ large enough.

To study the expected scenarios for the soft- and hard-core potentials, we consider two respective models for $V_{\mathrm{int}}(r)$: the hard-sphere potential of diameter $R$,
\begin{equation}
V_{\text{\textrm{hard}}}(r)=
\begin{cases}
\infty , & r<R \\
0, & r\geq R
\end{cases}
,  \label{Eq:V:HS}
\end{equation}
and its soft-sphere counterpart,
\begin{equation}
V_{\mathrm{soft}}(r)=
\begin{cases}
V_{0}, & r<R \\
0, & r\geq R
\end{cases}
,  \label{Eq:V:SS}
\end{equation}
with finite $V_{0}$ in the latter case.
By varying height $V_{0}$ of the soft-sphere potential one can change the value of the respective $s$-wave scattering length, which is
\begin{equation}
a_{s}=R[1-\tanh (kR)/(kR)],  \label{Eq:a:SS}
\end{equation}
where $k\equiv \sqrt{mV_{0}}/\hbar$ is the momentum associated with the height of the soft sphere.
For hard-sphere potential~(\ref{Eq:V:HS}), the effective $s$-wave scattering length is identical to the diameter of the sphere, $a_{s}=R$.
Note that the $s$-wave scattering length ~(\ref{Eq:V:HS}), corresponding to the soft-sphere potential, is smaller than the diameter, approaching it in the limit of $V_{0}\rightarrow \infty $.
As mentioned above, for dilute gases it is expected that $a_{s}$ is the single parameter which determines effects of the interparticle interactions \cite{Pitaevskii}.
Below, we compare the interaction potentials with the same $s$-wave scattering length $a_{s}$ but different diameters.

\section{The Monte Carlo method\label{sec:V}}

An efficient way to calculate the energy of a many-body system is to use the Monte Carlo technique.
We resort to the variational Monte Carlo method, which samples the probability distribution, $p=|\psi |^{2}$, for a known many-body wave function, $\psi$, allowing one to obtain the variational energy as a function of trial parameters, such as $\alpha$ and $\beta$ in Eq.~(\ref{Eq:f1}).
The standard Metropolis algorithm \cite{Metropolis} is used for the implementation of the method.

The many-body trial wave function is chosen as a product of single-particle terms, $f_{1}(r)$, and a pairwise product of two-particle Jastrow terms \cite{Jastrow}, $f_{2}(r)$:
\begin{equation}
\psi (\mathbf{r}_{1},\dots ,\mathbf{r}_{N})
= \prod\limits_{i=1}^{N}f_{1}(r_{i})\prod\limits_{i<j}^{N}f_{2}(|\mathbf{r}_{i}-\mathbf{r}_{j}|)  \label{Eq:f2}
\end{equation}
The single-particle factor is taken in the same form as the above single-particle ansatz, given by Eq.~(\ref{Eq:f1}), with parameter $\alpha$ controlling the degree of the particle's localization.
In particular, $\alpha \rightarrow 0$ corresponds to the limit of free delocalized particles, while $\alpha \rightarrow \infty$ implies the fall onto the center.
The Jastrow factor $f_{2}(r)$ in Eq.~(\ref{Eq:f2}) is chosen as a solution of the two-body scattering Schr\"{o}dinger equation.
In this way, the short-range physics in taken into account correctly, and interparticle correlations, which are important in the context of the metastability of a many-body system, are introduced.
For the hard-sphere potential, the two-body solution is given by
\begin{equation}
f_{2}^{\mathrm{(hard)}}(r)=
\begin{cases}
0, & r<R \\
1-R/r, & r\geq R
\end{cases}
.  \label{Eq:f2:HS}
\end{equation}
Note that, for distances larger than the hard-sphere's radius, solution $1-R/r$ is the same as for a pseudopotential with the respective $s$-wave scattering length, $a_{s}\equiv R$.

For the soft-sphere potential~(\ref{Eq:V:SS}), the two-body scattering solution is \cite{LL}
\begin{equation}
f_{2}^{\mathrm{(soft)}}(r)=
\begin{cases}
A\sinh (kr)/r, & r<R \\
1-a_{s}/r, & r\geq R
\end{cases}
\label{Eq:f2:SS}
\end{equation}
where the momentum scale is $k=\sqrt{mV_{0}}/\hbar $, see Eq.~(\ref{Eq:a:SS}), and constant $A$ is determined by the condition of the continuity of $f_{2}(r)$ at $r=R$.
Two hard spheres cannot overlap, hence the Jastrow factor (\ref{Eq:f2:SS}) vanishes at $r<R$, as written in Eq.~(\ref{Eq:f2:HS}).
On the other hand, for the soft spheres, there is a finite possibility of finding two particles at the same place, which allows the Jastrow factor to remain finite at $r=0$: $f_{2}^{\mathrm{(soft)}}(r=0)=Ak\approx (R-a_{s})k\exp (-kR)>0$, with $a_{s}<R$.
The possibility for two particles to overlap gets exponentially suppressed as the height of the soft sphere increases.
Two-body scattering solutions~(\ref{Eq:f2:HS}) and~(\ref{Eq:f2:SS}) coincide in the limit of $V_{0}\rightarrow \infty$.

\section{Numerical results for the many-body system \label{sec:VI}}

\subsection{The two-particle system \label{sec:N=2}}

\begin{figure}[tbp]
\includegraphics[width=0.4\columnwidth,angle=0]{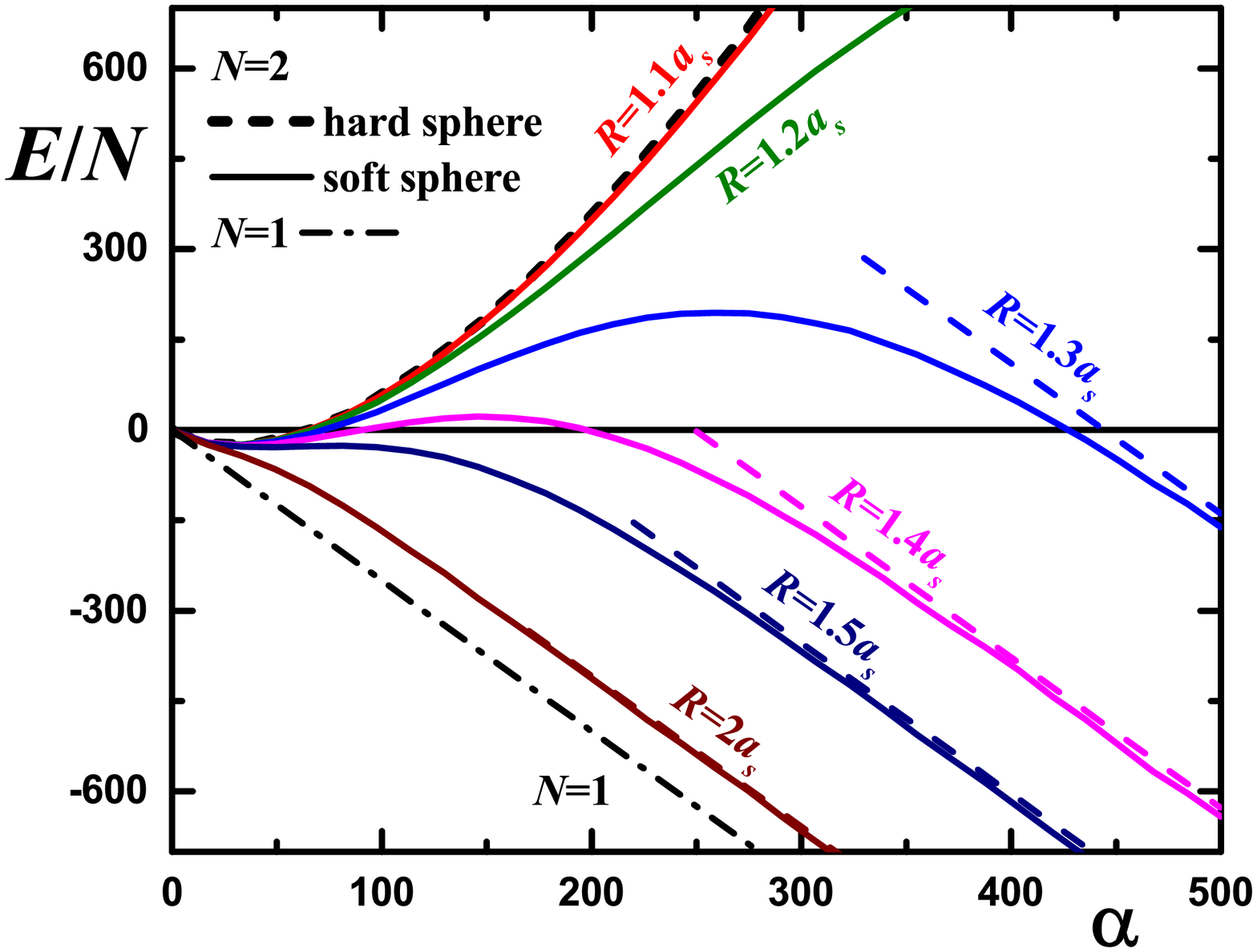}
\caption{(Color online) The variational energy per particle as a function of the inverse-Gaussian-width parameter, $\alpha$, for $U_{0}=1$, $a_{s}=0.1$ and $N=2$ particles, with the hard- and soft-sphere potentials.
Solid lines pertain to the soft-sphere potential with the diameter of the sphere, $R$, ranging from $1.1a_{s}$ to $2a_{s}$ (from the upper to the lower curve).
The dashed line corresponds to the hard-sphere potential ($R=a_{s}$).
The dash-dotted line depicts the single-particle energy given by Eq.~(\ref{Eq:E1}).
}
\label{FigEN2}
\end{figure}

First, we address the response of the two-particle system to compression.
To this end, we calculate, by means of the Monte Carlo method, the dependence of the variational energy on the density, which is controlled by localization parameter $\alpha $ in Eq.~(\ref{Eq:f1}). Figure~\ref{FigEN2} shows the variational energy for $U_{0}=1$ and $\beta=0$.
It is seen that pairs of particles interacting via the hard-sphere potential are always stable against the many-body collapse.
In the system with zero density, $\alpha=0$, the energy is also zero, $E=0$.
There is a minimum in the energy at a finite value of $\alpha$, meaning that the GS corresponds to a bound state ($E<0$) of a finite size.
For large values of $\alpha$ (i.e., for a small size of the localized system), the energy becomes large due to the strong interparticle repulsion.
A soft-sphere potential with a small size, $R=1.1a_{s}$, which, according to Eq.~(\ref{Eq:a:SS}), means large potential height $V_{0}$, follows behavior similar to that of the hard-sphere potential for the values of $\alpha $ shown in Fig.~\ref{FigEN2}.

For larger sizes of the soft sphere and smaller potential strengths, $V_{0}$, a second minimum in the variational energy becomes visible at finite values of $\alpha$ reported in Fig.~\ref{FigEN2}.
The absolute energy minimum stays at $\alpha\rightarrow \infty$, implying, as said above, that the GS corresponds to the fall onto the center (quantum collapse).
The gaseous state is, nevertheless, metastable if there is an energy barrier separating the two minima.
In the present case, we observe that a finite barrier is present at $U_{0}\lesssim 1.4$.
For a stronger attraction to the center, there is only one minimum, corresponding to the collapse.

Next, we aim to approximate the energy of a highly compressed state.
For a very strong localization, $\alpha \rightarrow \infty$, the bosonic particles stay on top of each other, hence the potential energy of the two-body interaction is well approximated by the height of the soft-sphere~(\ref{Eq:V:SS}), $E_{\mathrm{int}}=V_{0}$.
The two-body repulsion leads to an additional increase of the system's size, thus decreasing the kinetic energy per particle, $E_{\mathrm{kin}}/N$, and reducing the attraction to the center,  $E_{\mathrm{central}}/N$.
We find numerically that the net effect amounts to an increase the sum of both energies by $V_{0}$, against the one-particle value~(\ref{Eq:E1}), $E_{\mathrm{kin}}+E_{\mathrm{center}}=2E^{(1)}+V_{0}$.
As a result, the total energy of the two-particle system, $E^{(2)}$, can be approximated by
\begin{equation}
E^{(2)}=2E^{(1)}+2V_{0}.  \label{Eq:E2}
\end{equation}
The corresponding asymptotic behavior is shown in Fig.~\ref{FigEN2} by dashed lines.
In the limit of $\alpha \rightarrow \infty $, the energy of the two-particle system with the soft-sphere interaction is divergent and negative, corresponding to the quantum collapse.

\subsection{Many-body systems}

It has been demonstrated above that the energy barrier may be present between the gaseous and collapsed states for the two-particle system, $N=2$.
With the increase of $N$, the interparticle interactions improve the stability of the former state.

\begin{figure}[tbp]
\includegraphics[width=0.4\columnwidth, angle=0]{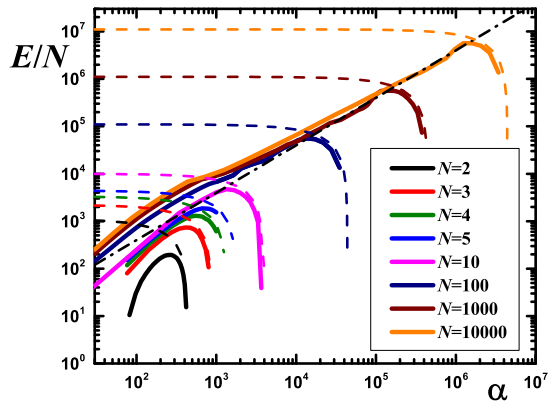}
\caption{(Color online) The energy per particle for the soft-sphere interaction potential as a function of the inverse-Gaussian-width parameter, $\alpha $, for $U_0=1$, $a_{s}=0.1$, $R=1.3a_{s}$ and the number of particles $N=2; 3; 4; 5; 10, 100, 1000, 10000$ (larger number of particles has larger value at the maximum). Solid lines, variational result; dashed lines, the asymptotic energy of the fully-collapsed state, as per Eq.~(\ref{Eq:EN}); dash-dotted line, typical energy associated with the Gaussian localization, given by Eq.~(\ref{Eq:E:C}). }
\label{Fig3}
\end{figure}

Figure~\ref{Fig3} shows the variation energy, calculated by the Monte Carlo method for a fixed radius of the soft sphere, $R=1.3a_{s}$, for a wide range of the number of particles, $N=2-10000$.
For localization parameter $\alpha=0$ the density vanishes and the energy is equal to zero.
For small values of $\alpha$, which corresponds to the weak localization, the energy might be negative (not visible on the log-log plot of Fig.~\ref{Fig3}).
As the localization gets tighter, the energy becomes positive, as the two-body interaction helps the system to resist the compression.
For very tight localization, $\alpha \rightarrow \infty$, the fall onto the center is observed with the energy diverging to minus infinity.

For large $N$, the energy calculated with ansatz~(\ref{Eq:f1}) does not immediately lead to the fully collapsed state.
The localization energy, proportional to $\hbar \omega$ of the harmonic trapping, may become a dominating term in the energy, while the system's size is still large enough, so that the fully-collapsed state is not realized.
The energy in the corresponding regime is numerically approximated as
\begin{equation}
E=NC\alpha   \label{Eq:E:C}
\end{equation}
with $C=4$, and is shown in Fig.~\ref{Fig3} by the dashed-dotted line.
For still tighter localization, it has been found that the following expression well approximates the energy of the system in the limit of the full collapse, when all particles overlap:
\begin{equation}
E=NE^{(1)}+N(N-1)V_{0}.  \label{Eq:EN}
\end{equation}
Similarly to the case of the two-body system [see Eq.~(\ref{Eq:E2})], the net effect of the interparticle interactions, revealed by our calculations, is two times larger than the potential energy, $E_{\mathrm{int}} = V_{0}N(N-1)/2$.
The asymptotic energy~(\ref{Eq:EN}) is shown in Fig.~\ref{Fig3} by dashed lines.

A clear conclusion is that the increase of the number of particles indeed causes strong rise of the potential barrier which stabilizes the energy minimum corresponding to the gaseous state.
This can be also concluded from Eq.~(\ref{Eq:EN}), where the contribution due to the repulsive interactions scales as $N^{2}$ for large $N$, while the term corresponding the attractive central potential scales as $N$.

\begin{figure}[tbp]
\includegraphics[width=0.4\columnwidth, angle=0]{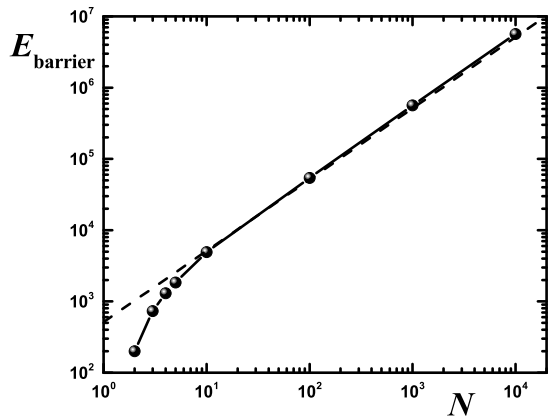}
\caption{The energy barrier between the state with $\alpha=0$ and $\alpha\to \infty$ as a function of the number of particles, $N$, for the data shown in Fig.~\ref{Fig3}.
The dashed line depicts the asymptotic approximation~(\ref{large}) for the large system.
}
\label{Fig4}
\end{figure}

The energy barrier between the state described by ansatz~(\ref{Eq:f1}) and the free state with zero energy,  $E_{\mathrm{barrier}}$, is estimated as the maximum value of the energy per particle (see Fig.~\ref{Fig3}), and is shown in Fig.~\ref{Fig4}.
For a large system size, the barrier can be approximated by comparing the two basic energy scales given by Eqs.~(\ref{Eq:E:C}) and~(\ref{Eq:EN}).
This resulting asymptotic approximation is
\begin{equation}
E_{\mathrm{barrier}}=6NV_{0}/(8U_{0}-3+2C),  \label{large}
\end{equation}
which is shown in Fig.~\ref{Fig4} by the dashed line.

\section{Conclusions\label{sec:VII}}

This work is stimulated by the recent results showing that, in the framework of the MFA (mean-field approximation), the quantum collapse (alias \textquotedblleft fall onto the center\textquotedblright) of bosons pulled by 3D attractive potential~(\ref{pot}) can be suppressed by the repulsive collisional nonlinearity in the ultracold bosonic gas, which is described by the GPE (Gross-Pitaevskii equation)~\cite{HS,HS2}.
Accordingly, the collapsing GS (ground state) is replaced by a regular one with a finite density.
Here, going beyond the framework of the MFA, we demonstrate that this prediction finds its counterpart in the many-body system governed by the same Hamiltonian: while the collapse is still possible in the system, the repulsive interactions create a non-singular metastable gaseous state, which is isolated from the collapsing one by a tall potential barrier.
The metastability quickly enhances with the increase of the number of bosons, $N$, starting from $N=2$, in agreement with the expectation that the MFA should be valid for $N$ large enough.
In the same time, the absolute stability threshold remains the same, in terms of Eq.~(\ref{pot}),  $U_{0}=1/8$, as for the single particle~\cite{LL}.
The results have been obtained through calculations of the variational energy, implemented by means of the Monte Carlo technique.

The analysis reported here may be naturally expanded in other directions.
In particular, it may be possible to construct stabilized states carrying the angular momentum, i.e., counterparts of vortical modes, predicted in Ref.~\cite{HS} with the help of a crude version of the MFA. Further, following the MFA results for the 2D geometry reported in Ref.~\cite{HS}, it may be interesting to consider the many-body system in the 2D setting.
The conclusion that only the quintic term may suppress the quantum collapse in that case, suggests that the stabilization of the 2D state may demand adding three-body interactions to the many-particle Hamiltonian.
Lastly, a challenging problem is to explore a similar possibility for fermions pulled to the center by potential~(\ref{pot}).
The simplest approach may then be to consider a balanced Fermi gas, with repulsive interactions between equal numbers of spin-up and spin-down particles~\cite{Yang,Sadhan}.

B.A.M. appreciates hospitality of ICFO -- Institut de Ci\`{e}ncies Fot\`{o}niques (Barcelona, Spain). G.E.A. acknowledges partial financial support from the MICINN (Spain) Grant No.~FIS2014-56257-C2-1-P.
The Barcelona Supercomputing Center (The Spanish National Supercomputing Center -- Centro Nacional de Supercomputaci\'{o}n) is acknowledged for the providing access to computational facilities.

\section{Appendix: The virial theorem \label{sec:Appendix}}

In this Appendix we derive the virial theorem for the pseudopotential interaction in presence of both the central attractive potential~(\ref{pot}) and harmonic-oscillator trapping.
For the sake of completeness we consider arbitrary dimension, $D=1,2,3$.

In the general form, the Hamiltonian is given the following sum
\begin{equation}
\hat{H}=\hat{E}_{\mathrm{kin}}+\hat{E}_{\mathrm{central}}+\hat{E}_{\mathrm{ho}}+\hat{E}_{\mathrm{int}},  \label{Eq:Hvirial}
\end{equation}
where
\begin{equation}
\hat{E}_{\mathrm{kin}}=-\frac{\hbar ^{2}}{2m}\sum\limits_{j=1}^{N}\nabla_{j}^{2}
\end{equation}
is the kinetic energy,
\begin{equation}
\hat{E}_{\mathrm{central}}=-\sum\limits_{j=1}^{N}\frac{U_{0}}{r_{j}^{2}}
\end{equation}
describes the attraction to the central potential,
\begin{equation}
\hat{E}_{\mathrm{ho}}=\frac{1}{2}m\omega ^{2}\sum\limits_{j=1}^{N}r_{j}^{2}
\end{equation}
is the potential energy of a the harmonic-oscillator confinement, and
\begin{equation}
\hat{E}_{\mathrm{int}}=\sum\limits_{j<k}^{N}g\delta (|\mathbf{r}_{j}-\mathbf{r}_{k}|)
\end{equation}
determines the interparticle interactions mediated by the pseudopotential.
Strictly speaking, in $\hat{E}_{\mathrm{int}}$ regularization of the $\delta$-functional interaction is needed in three dimensions, but it will not affect the scaling argument used here, therefore we disregard the regularization.

We now aim to check how the rescaling of particles' coordinates, $\mathbf{r}\rightarrow \xi \mathbf{r}$, affects averages of the different terms in the Hamiltonian.
The normalized GS wave function transforms according to
\begin{equation}
\psi (\mathbf{r}_{1},...,\mathbf{r}_{N},t)\rightarrow \frac{\psi (\xi \mathbf{r}_{1},...,\xi \mathbf{r}_{N})}{\sqrt{\int |\psi (\xi \mathbf{r}_{1},...,\xi \mathbf{r}_{N})|^{2}\;d\mathbf{r}_{1}...d\mathbf{r}_{N}}}
=\xi^{1/2}\;\psi (\xi \mathbf{r}_{1},...,\xi \mathbf{r}_{N}),
\end{equation}
while the energy becomes
\begin{equation}
E = \int \psi ^{\ast }\hat{H}\psi \;d\mathbf{r}_{1}...d\mathbf{r}_{N}
= \xi^{2}\langle \hat{E}_{\mathrm{kin}}\rangle +\xi ^{2}\langle \hat{E}_{\mathrm{cental}}\rangle +\xi ^{-2}\langle \hat{E}_{\mathrm{ho}}\rangle +\xi^{D}\langle \hat{E}_{\mathrm{int}}\rangle .
\end{equation}

The principle that the energy must be minimal for the GS imposes the following condition for a small variation around the ground-state solution with $\xi = 1$:
\begin{equation}
\left. \xi \frac{\delta E}{\delta \xi }\right\vert _{\xi =1}=0.
\end{equation}
As a result, we obtain the following relation (recall that $d$ is the spatial dimension):
\begin{equation}
2\langle \hat{E}_{\mathrm{kin}}\rangle +2\langle \hat{E}_{\mathrm{central}}\rangle -2\langle \hat{E}_{\mathrm{ho}}\rangle +D\langle \hat{E}_{\mathrm{int}}\rangle = 0,  \label{Eq:virial}
\end{equation}
which is the virial theorem for Hamiltonian~(\ref{Eq:Hvirial}).
In the absence of central potential~(\ref{pot}), Eq.~(\ref{Eq:virial}) amounts to the known result [for $D=3$, see Eq.~(44) in Ref.~\cite{BEC-RMP}].

\end{document}